\documentclass[twocolumn,showpacs,preprintnumbers,amsmath,amssymb,superscriptaddress,nofootinbib]{revtex4}

\usepackage{graphicx}
\usepackage{dcolumn}
\usepackage{bm}

\newcommand{\kf}{k_{\rm F}}
\newcommand{\vlowk}{V_{{\rm low}\,k}}

\begin{document}

\title{Non-empirical pairing energy functional in nuclear matter and finite nuclei}

\author{K.~Hebeler}
\email[E-mail:~]{hebeler@triumf.ca}
\affiliation{CEA, Centre de Saclay, IRFU/Service de Physique Nucl\'eaire, F-91191 Gif-sur-Yvette, France}
\affiliation{TRIUMF, 4004 Wesbrook Mall, Vancouver, BC, V6T 2A3, Canada}
\affiliation{ECT*, Strada delle Tabarelle 286, I-38050 Villazzano (Trento), Italy}
\author{T.~Duguet}
\email[E-mail:~]{thomas.duguet@cea.fr}
\affiliation{CEA, Centre de Saclay, IRFU/Service de Physique Nucl\'eaire, F-91191 Gif-sur-Yvette, France}
\affiliation{National Superconducting Cyclotron Laboratory and Department of Physics and Astronomy, Michigan State University, East Lansing, MI 48824, USA}
\author{T.~Lesinski}
\email[E-mail:~]{tlesinski@utk.edu}
\affiliation{Universit{\'e} de Lyon, F-69003 Lyon, France;
             Universit{\'e} Lyon 1, F-69622 Villeurbanne, France;
             CNRS/IN2P3; Institut de Physique Nucl{\'e}aire de Lyon}
\affiliation{Department of Physics and Astronomy, University of Tennessee, Knoxville, TN 37996, USA}
\affiliation{Physics Division, Oak Ridge National Laboratory, Oak Ridge, TN 37831, USA}
\author{A.~Schwenk}
\email[E-mail:~]{schwenk@triumf.ca}
\affiliation{TRIUMF, 4004 Wesbrook Mall, Vancouver, BC, V6T 2A3, Canada}

\begin{abstract}
We study $^{1}$S$_{0}$ pairing gaps in neutron and nuclear matter as
well as $T=1$ pairing in finite nuclei on the basis of microscopic two-nucleon interactions. Special attention is paid to the consistency of the pairing interaction and normal self-energy contributions. We find that pairing gaps obtained from low-momentum interactions depend only weakly on approximation schemes for the normal self-energy, required in present energy-density functional calculations, while pairing gaps from hard potentials are very sensitive to the effective-mass approximation scheme.
\end{abstract}


\maketitle

\section{Introduction}

Medium-mass and heavy nuclei can be studied systematically through nuclear energy-density-functional (EDF) calculations~\cite{bender03a}. Within a single-reference implementation, the minimization of the energy-density functional leads to solving Hartree-Fock-Bogoliubov (HFB) equations~\cite{Rin80aBxxx}. However, nuclear energy functionals accounting for correlated single-particle motions and superfluidity employed so far are of (semi-)empirical character~\cite{bender03a}. It is a central goal to construct {\it non-empirical} energy-density functionals connected to two- and many-nucleon interactions in free space~\cite{Duguet06} in view of the challenges posed by exotic nuclei with an unusually large ratio of neutrons over protons. The development of low-momentum interactions based on renormalization group (RG) methods~\cite{Bogner03,Bogner:2006vp} opens up such a possibility, as they enable technically simpler many-body approaches~\cite{Bogner05, Tolos:2007bh, Bogner:2009un}.

Such a long-term project to connect the nuclear EDF to underlying nuclear interactions has recently been initiated, first focusing on the part of the energy functional that drives pairing properties of nuclei~\cite{Duguet04,Duguet07,Lesinski08}. A difficulty is that a quantitative description of superfluidity in nuclear systems is a delicate task that a priori requires the treatment of complicated many-body processes. In fact, an on-going discussion concerns the impact of medium polarization effects, beyond the direct term of the nucleon-nucleon (NN) interaction, on pairing properties of finite nuclei. In Refs.~\cite{Barranco04,Pastore08}, two thirds of the observed neutron pairing gap was accounted for in $^{120}$Sn by using the Argonne $v_{14}$ NN potential~\cite{Wiringa:1984tg} as pairing interaction and combining this with the semi-empirical Skyrme functional SLy4 in the particle-hole channel. Adding induced interactions and self-energy effects due to the exchange of collective fluctuations between nucleons moving in time-reversed states, the missing one third was recovered~\cite{Terasaki02,Barranco04, Pastore08, Gori05}. In Refs.~\cite{Duguet07,Lesinski08}, however, neutron and proton pairing gaps were found to be consistent with experimental data over a large range of semi-magic nuclei when low-momentum NN interactions $\vlowk$ were used as pairing interaction, combined with the same Skyrme functional in the particle-hole channel. In this case, one therefore expects that neglected many-body forces and polarization effects result in a small net contribution to pairing gaps in (known) finite nuclei.\footnote{Of course, the neglected contributions do not have to be individually small.}

Before addressing the contribution of many-body forces and collective fluctuations to pairing gaps, the aim of the present work is to understand the qualitative and quantitative mismatch between the two sets of results published in Refs.~\cite{Duguet07,Lesinski08} and Refs.~\cite{Barranco04, Pastore08}, respectively, which both employ a free-space NN interaction as pairing force and the semi-empirical Skyrme functional SLy4 in the particle-hole channel. The only difference between the two calculations resides in the intrinsic resolution scale (in the RG sense~\cite{Bogner03, Bogner:2006vp}) of the Argonne $v_{14}$ and $\vlowk$~NN interactions, which both reproduce the relevant low-energy scattering phase shifts~\cite{Hebeler07}. Starting from this observation, one is led to focus on the coupling of the pairing interaction to the normal self-energy, in particular when the latter is approximated by a momentum-independent effective mass as is the case for present EDF calculations. Settling this issue requires fully microscopic calculations, where both the normal and anomalous self-energies are computed consistently from the NN interaction. To date, this is only possible for infinite nuclear matter (INM)~(see, for example, Ref.~\cite{Baldo90}). Of course, the results obtained in INM cannot be extrapolated straightforwardly to finite nuclei. Nevertheless, calculations of pairing gaps in INM provide a base-line for the schemes used in finite nuclei in Refs.~\cite{Barranco04,Pastore08,Duguet07,Lesinski08} and allow us to probe the sensitivity of pairing gaps to approximations of the normal self-energy.

We stress that our goal is not to perform the most involved calculations of pairing gaps in INM, for example, including induced interaction and associated self-energy effects~\cite{BabuBrown,Wambach1992ik,Schwenk:2002fq,cao06a}. Rather, we work at lowest order in the many-body expansion, with special attention (i) to the consistency between the normal and anomalous self-energies when using either a hard or low-momentum NN interaction and (ii) to the effects of neglecting the momentum dependence of the effective mass and quasiparticle strength when solving the gap equation.

This paper is organized as follows. In Section~\ref{label}, the many-body frameworks used to expand the normal self-energy and pairing gap are set up for NN interactions characterized by either a low- or high-momentum resolution scale (in the RG sense~\cite{Bogner03,Bogner:2006vp}). Procedures to average the momentum dependence of the normal self-energy are discussed. Section~\ref{results} presents results for the effective masses and pairing gaps in pure neutron matter and symmetric nuclear matter. The impact of the momentum averaging of the normal self-energy on pairing gaps is analyzed. In Section~\ref{fin_nuclei}, we discuss the consequences of our findings in INM on the computation of neutron and proton gaps in semi-magic nuclei. We conclude and give an outlook in Section \ref{conclusions}.

\section{Many-body framework}
\label{label}

\subsection{Hamiltonian}
\label{interaction}

The basic ingredient to microscopic calculations is the Hamiltonian that incorporates two- and many-nucleon interactions constrained by scattering experiments and few-body properties. In the present work, we neglect many-nucleon interactions, although it is important to characterize their impact on pairing properties~\cite{Baldo07}. Many-body forces may change the value of the gap, especially toward higher density, but are not expected to alter the conclusions of this paper.

The setup of a meaningful expansion scheme for nuclear many-body calculations depends on the choice of NN interaction \cite{Bogner05, Tolos:2007bh, Bogner:2009un}. We consider two schemes that are currently used in low-energy nuclear structure and reaction calculations. First, we work in a scheme which attempts to model the short-range parts of nuclear forces explicitly and is thus characterized by a large intrinsic resolution scale $\Lambda_{\text{hard}}$. This will be referred to as a ``hard'' NN interaction. Second, we consider low-momentum interactions with lower intrinsic resolution scale $\Lambda_{\text{soft}}$, which we refer to as ``soft'' NN interactions.

For both cases, we generate the interaction matrix elements starting from the Argonne $v_{18}$ potential~\cite{Wiringa95} by solving the symmetrized RG equation~\cite{Hebeler07,Bogner:2006vp} with an exponential regulator $f(k,\Lambda) = \exp[-(k/\Lambda)^{2n_{\rm{exp}}}]$ with $n_{\rm{exp}}=7$. The hard interaction is obtained by evolving to $\Lambda_{\text{hard}} = 6.0\,\rm{fm}^{-1}$. Using such a cutoff-scale instead of the initial Argonne $v_{18}$ potential allows one to reduce the numerical complexity of the calculations while maintaining the features of a hard potential. The soft interaction $V_{\rm{low}\,k}$ is obtained by further evolving the RG equation to a typical scale $\Lambda_{\text{soft}} = 1.8\,\rm{fm}^{-1}$~\cite{Bogner03, Bogner:2006vp}.

\subsection{Expansion scheme}
\label{expansion}

Starting from hard interactions, the short-range parts must be summed before the interaction can be used in many-body calculations. The traditional approach is based on the Brueckner $G$ matrix~\cite{goldstone57a} and on the reorganization of binding- or self-energy expansion schemes in terms of the number of hole lines entering the retained diagrams~\cite{day67a,Zuo:2001bd}. In addition, there are calculations for nuclear matter and finite nuclei based on a self-consistent in-medium $T$ matrix, which sums particle-particle as well as hole-hole ladders (see, for example, Ref.~\cite{Dickhoff:2004xx}). However, both expansions in terms of the $G$ or $T$ matrix are nonperturbative for hard interactions, because they do not decouple low and high momenta~\cite{Bogner05, Bogner:2009un}.

Starting from low-momentum interactions~\cite{Bogner03, Bogner:2006vp} the high-momentum modes are decoupled. This offers the possibility to use a perturbative expansion, as was shown explicitly for the particle-particle-channel contributions to the energy~\cite{Bogner05}. It remains to be checked that the expansion is perturbative in the particle-hole channels.

It is thus crucial to realize that the many-body expansion scheme differs depending on the intrinsic resolution scale characterizing the Hamiltonian. A trivial, but essential, implication is that only the {\it complete} resummation expressed in terms of the {\it full} Hamiltonian provides results that are independent of the expansion scheme, whereas results computed at a given order in the relevant expansion may differ depending on the scheme used. This underlines the necessity to specify the scheme employed and to perform consistent calculations, such as computing normal and superfluid self-energies at the same order in the relevant expansion scheme. In addition, due to the necessity to rearrange the expansion scheme depending on the resolution scale $\Lambda$, results obtained through truncated calculations cannot be expected to be cutoff independent.

Table~\ref{tab:exp_schemes} compares the set of diagrams taken into account to compute the normal and anomalous self-energies to first and second order in the two expansion schemes considered. In both cases, the first-order anomalous diagram is characterized by the use of the direct NN interaction as the (particle-particle-irreducible) kernel entering the gap equation. The first-order normal self-energy is provided by the Hartree-Fock (HF) diagram when starting from soft interactions, whereas it leads to the Brueckner-Hartree-Fock (BHF) approximation when employing hard potentials. The second-order diagrams correspond to a non-collective treatment of screening and vertex corrections as is done in Ref.~\cite{cao06a}.

The goal of this paper is to study $^1$S$_0$ pairing gaps in nuclear matter and finite nuclei at lowest order in the two expansion schemes, with special attention to the consistency of the pairing interaction and normal self-energy contributions. In particular, we focus on the effect of the normal self-energy and of effective-mass approximation schemes on the pairing gaps.

\subsection{Normal self-energy}
\label{normal}

\begin{table}[t]
\vspace{6pt}
\begin{tabular}{lll|lll}
\multicolumn{2}{c}{Soft NN interaction} & & & \multicolumn{2}{c}{Hard NN interaction} \\
\multicolumn{2}{c}{(low $\Lambda$)} & & & \multicolumn{2}{c}{(large $\Lambda$)} \\ \hline
\parbox[l][1.0cm][c]{0.8cm}{$\Sigma^{(1)}_{\text{soft}}=$} &\parbox[l][1.0cm][c]{0.8cm}{\includegraphics[width=0.8cm]{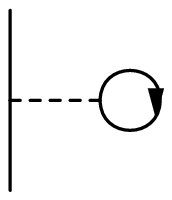}} & & &
\parbox[l][1.0cm][c]{0.8cm}{$\Sigma^{(1)}_{\text{hard}}=$} &\parbox[l][1.0cm][c]{0.8cm}{\includegraphics[width=0.8cm]{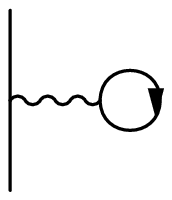}} \\
\parbox[l][1.0cm][c]{1.0cm}{$\Delta^{(1)}_{\text{soft}}=$} &\parbox[l][1.0cm][c]{0.8cm}{\includegraphics[width=0.8cm]{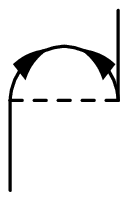}} & & &
\parbox[l][1.0cm][c]{1.0cm}{$\Delta^{(1)}_{\text{hard}}=$} &\parbox[l][1.0cm][c]{0.8cm}{\includegraphics[width=0.8cm]{eps/SHF_2.eps}} \\ \hline
\parbox[l][1.0cm][c]{0.8cm}{$\Sigma^{(2)}_{\text{soft}}=$} &
\parbox[l][1.0cm][c]{0.8cm}{\includegraphics[width=0.8cm]{eps/SHF_1.eps}} \parbox[l][1.0cm][c]{0.35cm}{$\:+$} \parbox[l][1.0cm][c]{0.8cm}{\includegraphics[width=0.8cm]{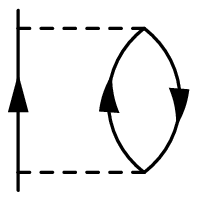}} \parbox[r][1.0cm][c]{0.35cm}{$\:+$} \parbox[l][1.0cm][c]{0.8cm}{\includegraphics[width=0.8cm]{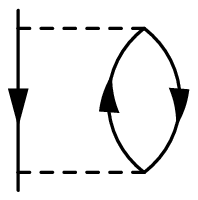}} & & &
\parbox[l][1.0cm][c]{0.8cm}{$\Sigma^{(2)}_{\text{hard}}=$} &
\parbox[l][1.0cm][c]{0.8cm}{\includegraphics[width=0.8cm]{eps/SBHF_1.eps}} \parbox[l][1.0cm][c]{0.35cm}{$\:+$} \parbox[l][1.0cm][c]{0.8cm}{\includegraphics[width=0.8cm]{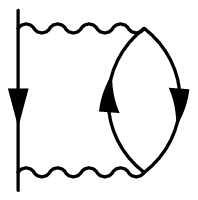}} \\
\parbox[l][1.0cm][c]{0.8cm}{$\Delta^{(2)}_{\text{soft}}=$} &
\parbox[l][1.0cm][c]{0.65cm}{\includegraphics[width=0.8cm]{eps/SHF_2.eps}} \parbox[l][1.0cm][c]{0.35cm}{$\:+$} \parbox[l][1.0cm][c]{0.8cm}{\includegraphics[width=0.8cm]{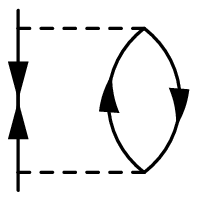}} & & &
\parbox[l][1.0cm][c]{0.8cm}{$\Delta^{(2)}_{\text{hard}}=$} &
\parbox[l][1.0cm][c]{0.65cm}{\includegraphics[width=0.8cm]{eps/SHF_2.eps}} \parbox[l][1.0cm][c]{0.35cm}{$\:+$} \parbox[l][1.0cm][c]{0.8cm}{\includegraphics[width=0.8cm]{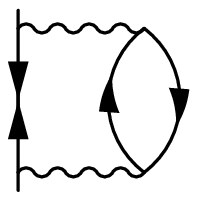}}
\end{tabular}
\caption{Expansion schemes for soft (left) and hard (right) interactions to first (up) and second (down) order. The dashed line denotes the free-space NN interaction and the wiggly line the $G$ matrix. Diagrams with more than one anomalous propagator are not shown.}
\label{tab:exp_schemes}
\end{table}
At lowest order, the normal self-energy takes the form
\begin{align}
\Sigma^{(1) \tau}_{\text{soft}} (p_{1})  &= \sum_{\mathbf{p}_{2},\tau',\sigma'} \hspace{-0.2cm} n^{\tau'} \hspace{-0.07cm} (p_{2}) \left< \, \mathbf{k} \, \right| V^{\tau \tau'} \left| \, \mathbf{k} \, \right>\, ,  \label{eq:sigma0} \\
\Sigma^{(1) \tau}_{\text{hard}}{} (p_{1}, \omega)  &= \sum_{\mathbf{p}_{2},\tau', \sigma'} \hspace{-0.2cm} n^{\tau'} \hspace{-0.07cm} (p_{2}) \left< \, \mathbf{k} \, \right| G^{\tau \tau'} (P, \omega \hspace{-0.05cm} + \hspace{-0.05cm} \varepsilon^{\tau'} \hspace{-0.07cm} (p_{2})) \left| \, \mathbf{k} \nonumber \, \right>\, , \\[-0.4cm] 
\label{eq:sigma}
\end{align}
with the notation $p\equiv|\mathbf{p}|$ and where the occupation function $n^{\tau}(p)\equiv\theta(\kf^{\tau}-|\mathbf{p}|)$ is taken as a step function. The relative and center-of-mass momenta are defined by $\mathbf{k} \equiv (\mathbf{p}_{1} - \mathbf{p}_{2})/2$ and $\mathbf{P} \equiv \mathbf{p}_{1} + \mathbf{p}_{2}$, respectively. The index $\tau$ characterizes the single-particle isospin projection. Using normal-state Fermi-Dirac distributions functions rather than BCS-like occupation numbers is a satisfactory approximation in INM around saturation density where the pairing gap $\Delta^{\tau}$ is small compared to the Fermi energy $\varepsilon^{\tau}_{\text{F}} \equiv (\kf^{\tau})^2/(2m)$~\cite{PhysRevC.59.2927}.

The Brueckner $G$ matrix is calculated through a partial-wave expansion (in units of $\hbar = c = 1$)
\begin{widetext}
\begin{equation}
\left< k' \right| G^{\tau \tau'}_{l l' S J} (P, \omega) \left| k \right>
= \left< k' \right| V^{\tau \tau'}_{l l' S J} \left| k \right> + \frac{2}{\pi} \int{q}^{\, 2} dq \, \sum_{\widetilde{l}} \left< k' \right| V^{\tau \tau'}_{l \widetilde{l} S J} \left| q \right> \frac{\big<Q^{\tau \tau'}\hspace{-0.07cm} (P,q)\big>}{\omega - \left< \varepsilon^{\tau \tau'}\hspace{-0.07cm} (P,q) \right> + i \delta}
\left< q \right| G^{\tau \tau'}_{\widetilde{l} l' S J} (P, \omega) \left| k \right> \, ,
\label{eq:Gmatr}
\end{equation}
\end{widetext}
where the angular-averaged Pauli-blocking operator and two-particle-state energies are defined as
\begin{align}
\big< Q^{\tau \tau'}\hspace{-0.07cm} (P,k) \big> &\equiv \frac{1}{2} \int d \cos \theta_{{\bf{P}\bf{k}}} \Big[1-n^{\tau}(p_{1})\Big]\Big[1-n^{\tau'} \hspace{-0.07cm} (p_{2})\Big]\, , \nonumber \\
\big<\varepsilon^{\tau \tau'} \hspace{-0.07cm} (P,k) \big> &\equiv \frac{1}{2} \int d \cos \theta_{{\bf{P}\bf{k}}} \left[ \varepsilon^{\tau} (p_{1}) + \varepsilon^{\tau'}\hspace{-0.07cm}(p_{2}) \right]\, . \nonumber
\end{align}
The on-shell single-particle energy $\varepsilon^{\tau} (p)$ entering Eqs.~(\ref{eq:sigma}) and~(\ref{eq:Gmatr}) is obtained through
\begin{equation}
\varepsilon^{\tau} (p) \equiv \frac{p^2}{2m} + \text{Re} \, \Sigma^{(1) \tau}_{\Lambda} (p, \varepsilon^{\tau} (p))\, .
\label{eq:epsilon}
\end{equation}
Equations~(\ref{eq:sigma}),~(\ref{eq:Gmatr}) and~(\ref{eq:epsilon}) are solved self-consistently when using hard interactions. For soft interactions, the system reduces to the direct evaluation of Eqs.~(\ref{eq:sigma0}) and~(\ref{eq:epsilon}) thanks to the energy independence of $\Sigma^{(1) \tau}_{\text{soft}}$.

We note that these equations are valid for asymmetric nuclear matter, although we restrict ourselves to pure neutron matter (PNM) and symmetric nuclear matter (SNM) in the present work. Because we focus on neutron-neutron pairing, only neutron self-energies, $\tau=n$, are eventually needed. Thus, the short-hand notations
$\varepsilon (p) \equiv \varepsilon^{n} (p), \Sigma^{\rm{HF}} (p) \equiv \Sigma^{(1) n}_{\text{soft}} (p)$ and $ \Sigma^{\rm{BHF}} (p,\omega) \equiv \Sigma^{(1) n}_{\text{hard}} (p,\omega)$ are used, along with corresponding notations for effective masses, quasiparticle strength, pairing gaps and Fermi momenta introduced below.

\subsection{Effective-mass approximation}
\label{effectivemass}

\subsubsection{Definitions}
\label{effectivemassdef}

A focus of the present work is to study effective-mass approximation schemes on pairing gaps. The momentum-dependent effective mass $m^*_{\tau} (p, \kf^{\tau'})$ is defined by

\begin{equation}
\frac{d \varepsilon^{\tau} (p)}{d p} \equiv \frac{p}{m^*_{\tau} (p, \kf^{\tau'})}\, , \label{totaleffectivemass}
\end{equation}
where $\kf^{\tau'}$ denotes the dependence on both Fermi momenta. This total effective mass can be separated into the product of the $k$-mass and the $e$-mass defined by~\cite{sartor80a}
\begin{align}
\frac{m^{*}_{\tau, k} (p,\kf^{\tau'})}{m} &\equiv \left[ 1 + \frac{m}{p} \left. \frac{\partial \text{Re} \, \Sigma^{\tau}_{\Lambda} (p, \omega)}{\partial p} \right|_{\omega = \varepsilon^{\tau} (p)}\right]^{-1} , \\
\frac{m^{*}_{\tau, e} (p,\kf^{\tau'})}{m} &\equiv 1 - \left. \frac{\partial \text{Re} \, \Sigma^{\tau}_{\Lambda} (p, \omega)}{\partial \omega} \right|_{\omega = \varepsilon^{\tau} (p)} .
\end{align}

The $k$-mass relates to the spatial non-locality of the normal self-energy, whereas the $e$-mass characterizes the dynamical correlations associated with the energy dependence. The $e$-mass can also be expressed in terms of the quasiparticle strength, or $Z$-factor,
\begin{equation}
Z_{\tau} (p,\kf^{\tau'}) \equiv \frac{m}{m^{*}_{\tau, e}(p,\kf^{\tau'})}\, ,
\end{equation}
which quantifies the quasiparticle part of the one-body Green's function and occurs in the pole approximation discussed in the following.

\subsubsection{Momentum-independent approximation}
\label{effectivemassapprox}

At this point, the introduction of the effective mass is essentially a matter of definition. The real purpose is usually to neglect its momentum dependence in order to recover a (density-dependent) quadratic dispersion relation for $\varepsilon^{\tau} (p)$. In the present work, it is motivated by the need to make the connection with the Skyrme functional whose self-energy can at best be related to a momentum-independent approximation of the microscopically-obtained effective mass.

Hence we have to reduce the momentum dependence of $m^*_{\tau} (p, \kf^{\tau'})$ and $Z_{\tau}(p, \kf^{\tau'})$ to a dependence on $\kf^{\tau}$. Obviously there are different ways to do so. In order to probe the sensitivity of observables to the particular scheme used, we consider two approximations. The first, which we denote as the point-evaluation (``$pe$'') approximation, is standard and consists of taking the value at the Fermi momentum, with $X_{\tau}=m^*_{\tau}$ or $Z_{\tau}$,
\begin{equation}
X_{pe} (\kf^{\tau'}) \equiv X_{\tau} (p=\kf^{\tau},\kf^{\tau'})\, .
\label{eq:pe}
\end{equation}
The second method proceeds through an averaging (``$av$'') of the momentum dependence over the Fermi surface,
\begin{equation}
X_{av} (\kf^{\tau'}) \equiv \frac{\int \hspace{-0.05cm} f(q,\Lambda) \, q^2 dq \, X_{\tau}(q,\kf^{\tau'}) \, \bar{u}_q^{\tau} \, \bar{v}_q^{\tau}}{\int \hspace{-0.05cm} f(q,\Lambda) \, q^2 dq \, \bar{u}_q^{\tau} \, \bar{v}_q^{\tau}}\, ,
\label{eq:av}
\end{equation}
where $\bar{u}_q^{\tau} \, \bar{v}_q^{\tau} \equiv \bar{\Delta} / (2 \sqrt{(\xi_0^{\tau} (q))^2 + \bar{\Delta}^2})$ denotes the BCS pair occupation function, which is peaked at the Fermi surface. The free single-particle spectrum $\xi_0^{\tau} (p)=(p^{2}- \kf^{\tau}{}^2)/(2m)$ and a typical width of $\bar{\Delta} = 2.0\,\rm{MeV}$ are used for simplicity in Eq.~(\ref{eq:av}).

\subsection{Anomalous self-energy}
\label{gaps}

After discussing the computation of the normal part of the self-energy together with its effective-mass approximation, we now turn to the anomalous self-energy that leads to pairing gaps. We are interested in neutron-neutron pairing at sub-saturation densities $\kf \approx 0.8 - 1.4\,{\rm fm}^{-1}$ in PNM and SNM. Neutron superfluidity in this density range in INM is in the $^{1}\text{S}_{0}$ channel~\cite{Dean03, Hebeler07}. In this work the pairing kernel is restricted to the direct NN interaction and therefore we keep only the $^{1}\text{S}_{0}$ partial wave as pairing interaction.

The gap equation is solved within the pole approximation, which provides a good approximation to the solution of the full off-shell gap equation when the momentum dependence of the effective mass and of the $Z$-factor is taken into account~\cite{Bozek03}. Furthermore, and as already mentioned, the normal self-energy and quasiparticle strength are computed in the normal state, which is valid for the density range considered where $\Delta/\varepsilon_{\text{F}} \ll 1$. In summary, the neutron anomalous self-energy and thus the gap $\Delta(p)$ is the solution of \cite{Baldo00b}
\begin{equation}
\Delta (p) = - \frac{1}{\pi} \int dq \, q^2 \: \frac{\big< p | V_{^1\rm{S}_0}^{nn} | q \big> \, Z (q) \, \Delta (q)}{\sqrt{ \widetilde{\xi}^2 (q)  + {\Delta^2 (q)}}}\, , \label{gap1}
\end{equation}
with
\begin{equation}
\widetilde{\xi} (p) \equiv \frac{p^2}{2m} - \mu + \frac{1}{2} \Big[ \text{Re} \, \Sigma_{\Lambda}(p, \varepsilon (p)) + \text{Re} \, \Sigma_{\Lambda}(p, 2 \mu - \varepsilon (p)) \Big] \, , \label{symself} \nonumber 
\end{equation}
and the chemical potential is defined by $\mu \equiv \varepsilon (\kf)$, so that it includes the normal self-energy shift with respect to the free Fermi energy. The chemical potential can also be calculated self-consistently to account for the effect of pairing correlations, but we have checked that this has very little effect in the density range of interest.

It is important to realize that for $Z (q) \neq 1$ the physical gap $\widehat{\Delta} (p = \kf)$ in the excitation spectrum of the system is given by~\cite{PhysRevC.62.054316}
\begin{equation}
\widehat{\Delta} (p) \equiv Z (p) \, \Delta (p)\, .\label{gap2}
\end{equation}
Linearizing additionally $\Sigma_{\Lambda}(p, \omega)$ in energy around $\mu$ leads to the BCS-type equation
\begin{equation}
\widehat{\Delta}_i (p) = - \frac{1}{\pi} \int dq \, q^2 \frac{Z_i(p) \, \big< p | V_{^1\rm{S}_0}^{nn} | q \big> \, Z_i(q) \, \widehat{\Delta}_i (q)}{\sqrt{ {\xi_i^2 (q)} + {\widehat{\Delta}^2_i (q)}}}\, ,\:\:\:
\label{eq:gap}
\end{equation}
where $\xi_i (p) = \varepsilon_i (p) - \mu$.

The index $i$ in Eq.~(\ref{eq:gap}) labels different cases considered in this paper regarding the choice of single-particle energy and quasiparticle strength. We define three classes depending on whether one starts from a soft or a hard NN interaction, see Table~\ref{tab:eps_classes}. For each class, one goal is to compare the gaps obtained from strictly solving Eq.~(\ref{eq:gap}) to those obtained using further approximations, for example neglecting the momentum dependence of the effective mass and of the $Z$-factor. The third class ($\Lambda_{\text{hard}}^1$) defined in Table~\ref{tab:eps_classes} is not consistent in the sense that we only keep the $k$-mass of the total BHF effective mass and neglect all $e$-mass and $Z$-factor effects in the gap equation\footnote{Once the $k$-mass is extracted from the total effective mass, the corresponding self-energy is recovered through the integral given in Table~\ref{tab:eps_classes}.}. This can be considered as an intermediate case between classes $\Lambda_{\text{soft}}^1$ and $\Lambda_{\text{hard}}^Z$.

\begin{table}
\begin{tabular}{|ll|l|}
\hline
\multicolumn{3}{|c|}{\parbox[c][0.6cm][c]{0.5cm}{$\text{(i)} \, \, \Lambda_{\text{soft}}^1$}} \\
\hline
\parbox[c][0.7cm][c]{0cm}{} & $\xi_1(p) = \varepsilon^{\rm{HF}} (p) - \mu$ & $Z_1 (p)=1$ \\
\parbox[c][0.7cm][c]{0cm}{} & $\xi_{2/3}(p) = (p^2 - \kf^2)/(2  m^{* \rm{HF}}_{pe/av} (\kf))$ & $Z_{2/3} (p)=1$ \\
\hline
\hline
\multicolumn{3}{|c|}{\parbox[c][0.6cm][c]{0.5cm}{$\text{(ii)} \, \, \Lambda_{\text{hard}}^Z$}} \\
\hline
\parbox[c][0.7cm][c]{0cm}{} & $\xi_1(p) = \varepsilon^{\rm{BHF}} (p) - \mu$ & $Z_1 (p) = Z^{\rm{BHF}}(p)$ \\
\parbox[c][0.7cm][c]{0cm}{} & $\xi_{2/3} (p) = (p^2 - \kf^2)/(2  m^{* \rm{BHF}}_{pe/av} (\kf) )$ & $Z_{2/3} (p)=Z^{\rm{BHF}}_{pe/av} (\kf)$ \\
\hline
\hline
\multicolumn{3}{|c|}{\parbox[c][0.6cm][c]{0.5cm}{$\text{(iii)} \, \, \Lambda_{\text{hard}}^1$}} \\
\hline
\parbox[c][0.7cm][c]{0cm}{} & $\xi_1 (p) = \int_{\kf}^{p} dq \: q/m_k^{* \rm{BHF}} (q)$ & $Z_1 (p)=1$ \\
\parbox[c][0.7cm][c]{0cm}{} & $\xi_{2/3} (p) = (p^2 - \kf^2)/(2  m^{* \rm{BHF}}_{k,pe/av} (\kf) )$ & $Z_{2/3} (p) = 1$ \\
\hline
\end{tabular}
\caption{Three classes of calculations: (i) first order in a soft NN interaction, (ii) first order in a hard NN interaction, (iii) first order in a hard NN interaction but neglecting all effects related to the energy dependence of the normal self-energy. Each class contains a calculation of reference retaining the full momentum dependence of the normal self-energy ($i=1$) and the $pe/av$ schemes ($i=2, 3$).}
\label{tab:eps_classes}
\end{table}

\section{Infinite nuclear matter results}
\label{results}

\subsection{Effective mass}
\label{results_effmass}

\subsubsection{Momentum dependence}
\label{results_effmassmomentumdep}

\begin{figure}
\includegraphics[clip]{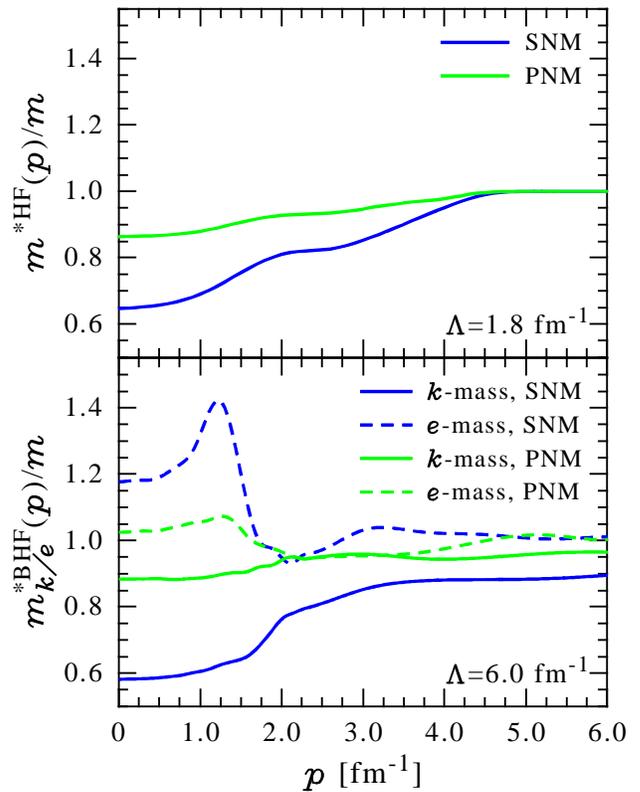}
\caption{(Color online) Momentum dependence of the HF effective $k$-mass for $\Lambda_{\text{soft}}=1.8\,\rm{fm}^{-1}$ (upper panel) and of the BHF effective masses for $\Lambda_{\text{hard}}=6.0\,\rm{fm}^{-1}$ (lower panel) in PNM and SNM at $\kf=1.2\,\rm{fm}^{-1}$. Whereas the HF approximation only provides a $k$-mass, the hole-line expansion already generates an $e$-mass at leading order.}
\label{fig:mk}
\end{figure}
Figure~\ref{fig:mk} shows the momentum dependence of the effective masses calculated at lowest order for soft (upper panel) and hard (lower panel) interactions, at a representative density of $\kf=1.2\,\rm{fm}^{-1}$ in PNM and SNM.

The HF calculation using the soft interaction only generates a $k$-mass that is smaller than the bare mass and displays a smooth momentum dependence. The HF $k$-mass is smaller in SNM than in PNM due to the stronger in-medium effects, notably brought by the proton-neutron (tensor) interaction.

In the BHF calculation already at leading order a $k$-mass and an $e$-mass are generated. The $e$-mass is given by the energy dependence of $\rm{Re}\, \Sigma^{(1)}_{\text{hard}}{} (p, \omega)$ and displays a typical enhancement around the Fermi momentum associated with the increased probability of virtually occupy two-particle--one-hole configurations. This effect comes at second order in the perturbative expansion for soft interactions (see Table~\ref{tab:exp_schemes}). The BHF $k$-mass is similar to, but slightly smaller than the HF $k$-mass shown on the upper panel. The total BHF effective mass is the product of the $k$-mass and the $e$-mass and is thus larger than the $k$-mass for all densities, in addition to carrying the typical enhancement of the $e$-mass around the Fermi momentum. We have also checked that BHF masses and $Z$-factors do not change significantly by increasing the cutoff beyond $\Lambda=6.0\,\rm{fm}^{-1}$ in SNM and PNM.

\subsubsection{Averaged momentum dependence}
\label{results_effmassavermomentumdep}

\begin{figure}
\includegraphics{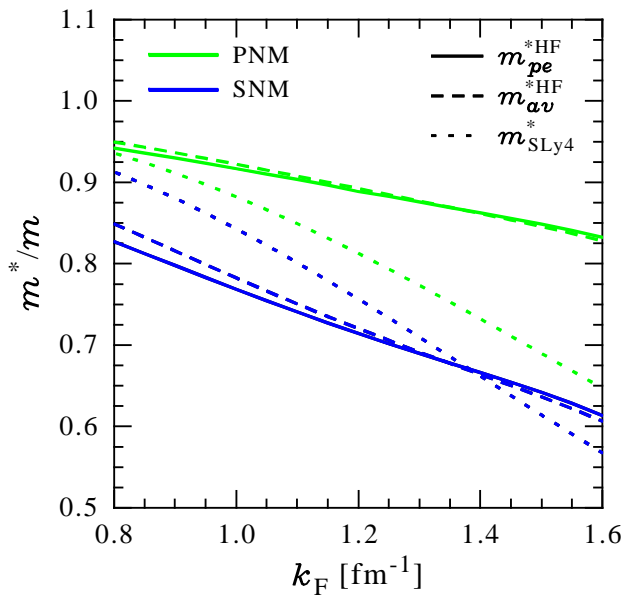}
\caption{(Color online) Momentum-independent effective masses $m^{* \, \rm{HF}}_{pe}(\kf)$ and $m^{* \, \rm{HF}}_{av}(\kf)$ obtained from the soft interaction in PNM and SNM. For comparison we show the effective masses $m^*_{\text{SLy4}}$ of the Skyrme SLy4 parameterization.}
\label{fig:mkf_HF_LSy4}
\end{figure}

Figure~\ref{fig:mkf_HF_LSy4} shows the momentum-independent effective masses $m^{* \, \rm{HF}}_{pe}(\kf)$ and $m^{* \, \rm{HF}}_{av}(\kf)$ obtained by applying the two averaging schemes introduced in Section~\ref{effectivemassapprox} for the soft interaction. A key result obtained for PNM and SNM is that the two schemes lead to essentially identical results. This is due to the mild momentum dependence of the HF $k$-mass obtained from the soft interaction and indicates that reducing such a momentum dependence may be a tractable approximation when solving the gap equation. Figure~\ref{fig:mkf_HF_LSy4} also compares the microscopically-calculated $m^{* \rm{HF}}(\kf)$ to the density-dependent effective mass $m^*_{\text{SLy4}}(\kf)$ of the SLy4 Skyrme functional parameterization. Although the functional dependence is not fully captured by $m^*_{\text{SLy4}}(\kf)$, the microscopic result in SNM at saturation density $\kf = 1.35\,\rm{fm}^{-1}$ is well reproduced. In PNM, however, the neutron effective mass $m^*_{\text{SLy4}}(\kf)$ underestimates the microscopic predictions significantly. This reflects the known deficiency in the isovector effective mass of SLy4~\cite{lesinski06a}.

\begin{figure}
\includegraphics{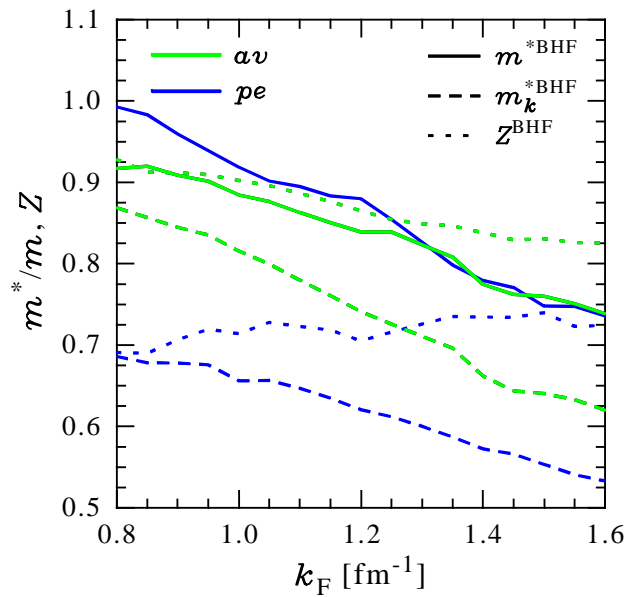}
\caption{(Color online) Momentum-independent effective masses and $Z$-factors obtained from the hard interaction in the BHF scheme in SNM. $m^{* \rm{BHF}}_{k} (\kf)$ and $Z^{\rm{BHF}} (\kf)$ depend strongly on the averaging scheme, whereas the total mass is less sensitive to this.}
\label{fig:mkf_SNM_BHF}
\end{figure}

In Fig.~\ref{fig:mkf_SNM_BHF}, we present the momentum-independent total effective mass, $k$-mass, and $Z$-factor obtained in SNM applying the two averaging schemes in the case of the hard interaction. Due to the more pronounced momentum dependence of the effective masses and the larger averaging region set by the regulator $f(q,\Lambda=6.0\,\rm{fm}^{-1})$ in Eq.~(\ref{eq:av}), the point-evaluated and averaged values differ substantially. This difference is also much larger in SNM than in PNM as self-energy effects are larger in SNM. Due to a compensation effect between the $k$-mass and the $e$-mass, the total effective mass happens to be relatively insensitive to the averaging scheme used. Nevertheless, as the gap equation depends on both the total effective mass and $Z$-factor (see Eq.~(\ref{eq:gap})), the present results indicate that it may be unreliable to average the momentum dependence of $m^{* \, \rm{BHF}} (p, \kf)$ and $Z (p, \kf)$ in this case. The same conclusion can be anticipated when solving the gap equation within the frame of the third class of Table~\ref{tab:eps_classes}, that is when the $e$-mass and $Z$-factor are neglected.

\subsection{Pairing gaps}
\label{results_gaps}

\begin{figure}[t]
\includegraphics[clip]{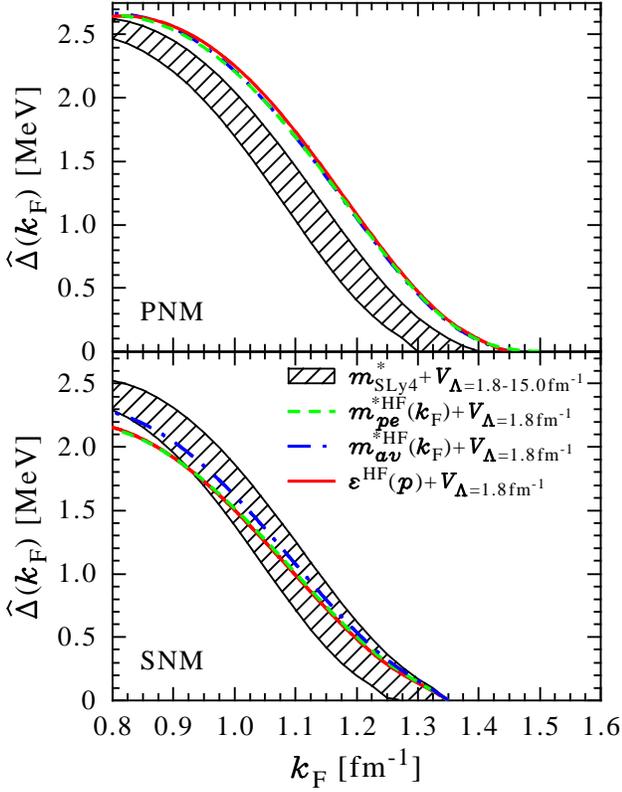}
\caption{(Color online) Neutron $^1$S$_0$ pairing gaps in PNM and SNM obtained using the soft interaction with $\Lambda_{\text{soft}}=1.8\,\rm{fm}^{-1}$ as the pairing interaction and the HF approximation for the normal self-energy. Results are shown for the three cases of class $\Lambda^1_{\rm{soft}}$ (see Table~\ref{tab:eps_classes}).}
\label{fig:gap_soft}
\end{figure}
\begin{figure}[t]
\includegraphics[clip]{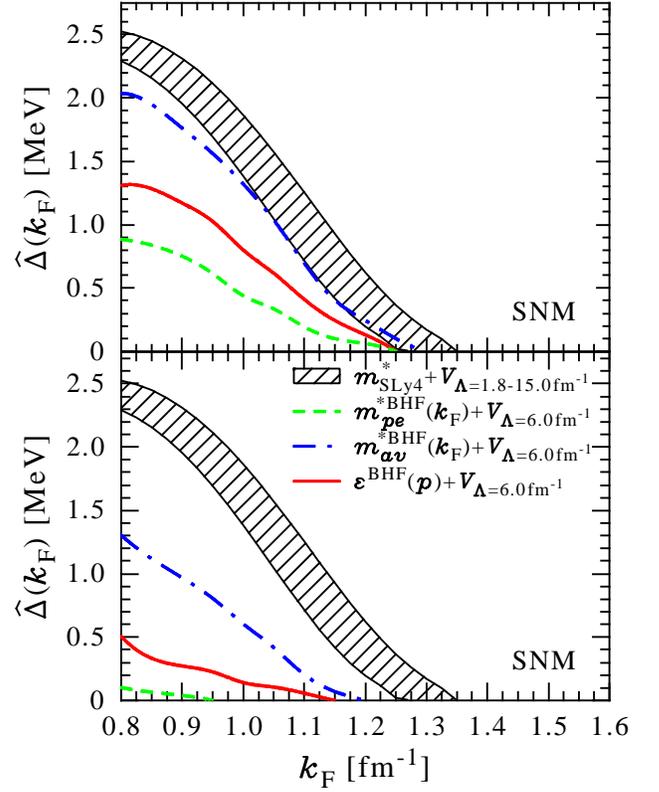}
\caption{(Color online) Neutron $^1$S$_0$ pairing gaps in SNM obtained using the hard interaction with $\Lambda_{\text{hard}}=6.0\,\rm{fm}^{-1}$ as the pairing interaction and the BHF approximation for the normal self-energy. For the results in the upper panel, only the $k$-mass effects are taken into account according to class $\Lambda_{\text{hard}}^1$ of Table~\ref{tab:eps_classes}. The lower panel shows the results of the class $\Lambda_{\text{hard}}^Z$ including $e$-mass and $Z$-factor effects.}
\label{fig:gap_hard}
\end{figure}

\subsubsection{Soft interaction}
\label{results_gaps_soft}

Figure~\ref{fig:gap_soft} shows our results for the pairing gaps at the Fermi surface $\widehat{\Delta} (p=\kf)$ as a function of the Fermi momentum $\kf$ in PNM and SNM for the $\Lambda_{\text{soft}}^1$ class defined in Table~\ref{tab:eps_classes}. The band represents the pairing gaps obtained using the SLy4 effective mass and varying the cutoff in the NN interaction over a wide cutoff range. This band corresponds to the variation of the pairing gaps obtained in finite nuclei~\cite{lesinski08b} by varying the resolution scale of the pairing kernel. The upper limit of the band corresponds to the low cutoff $\Lambda = 1.8\,\rm{fm}^{-1}$. This is the result to be compared to the microscopic calculations discussed in this subsection. The lower limit of the band corresponds to the hard cutoff $\Lambda = 15.0\,\rm{fm}^{-1}$ and will be relevant to the next subsection.

As shown in the upper panel of Fig.~\ref{fig:gap_soft}, the pairing gaps obtained from the $\Lambda^1_{\text{soft}}$ class are essentially indistinguishable. We find
\begin{equation}
\widehat{\Delta} \big[\varepsilon^{\rm{HF}} (k) \big] \approx \widehat{\Delta} \big[ m_{pe}^{* \rm{HF}} (\kf) \big] \approx \widehat{\Delta} \big[ m_{av}^{* \rm{HF}} (\kf) \big] \, ,
\label{eq:gaps_HF}
\end{equation}
to an excellent approximation in PNM. As expected from Fig.~\ref{fig:mkf_HF_LSy4}, the momentum averaging of the normal self-energy $\varepsilon^{\rm{HF}} (p) \rightarrow (p^2 - \kf^2)/(2  m^{* \rm{HF}} (\kf))$ is well justified in PNM and has essentially no impact on the computed gaps. Due to the wrong isovector dependence of $m^*_{\rm{SLy4}}$ compared to $m^{* \rm{HF}} (\kf)$  (see Fig.~\ref{fig:mkf_HF_LSy4}), the microscopic gaps are larger than the upper limit of the band.

In SNM the pairing gaps are also insensitive to the effective-mass approximation scheme, in particular over the range $\kf \approx 1.0-1.4\,\rm{fm}^{-1}$. In the density region $\kf \approx 1.2-1.4\,\rm{fm}^{-1}$, $m^*_{\rm{SLy4}}$ reproduces well the effective mass obtained from the soft NN interaction (see Fig.~\ref{fig:mkf_HF_LSy4}) and therefore the microscopic gaps are close to the upper limit of the band. At lower densities $\kf < 1.2\,\rm{fm}^{-1}$, $m^*_{\rm{SLy4}}$ is larger than the calculated effective mass and therefore we find pairing gaps that are smaller than the upper limit of the band.

In addition, we practically find cutoff independence of the pairing gaps in SNM and PNM for soft cutoffs $\Lambda \approx 1.8-3.0\,\rm{fm}^{-1}$. This approach therefore provides a tractable lowest-order starting point with respect to appropriate\footnote{One must consider variations of $\Lambda$ such that the perturbative expansion remains valid. Weinberg eigenvalues demonstrate that this is the case for $\Lambda \lesssim 3.0\,\rm{fm}^{-1}$~\cite{Bogner:2006tw}.} variations of the renormalization scale. These results also complement the cutoff independence of pairing gaps obtained in INM using a free single-particle spectrum~\cite{Hebeler07} and in finite nuclei~\cite{lesinski08b} over the same cutoff range.

\subsubsection{Hard interaction}
\label{results_gaps_hard}
Figure~\ref{fig:gap_hard} shows the pairing gaps $\widehat{\Delta} (p=\kf)$ computed in SNM according to the two classes $\Lambda_{\text{hard}}^{1}$ and $\Lambda_{\text{hard}}^{Z}$ defined in Table~\ref{tab:eps_classes}. The upper panel shows the results taking only the $k$-mass contributions into account ($\Lambda_{\text{hard}}^{1}$), whereas the lower panel includes also the $e$-mass and $Z$-factor effects ($\Lambda_{\text{hard}}^{Z}$) generated at first order in the hole-line expansion.

Comparing the two different classes, we find the typical systematic reduction of the gaps due to the decreased spectral strength of the quasiparticle propagator that wins over the increased density of states characterizing the $e$-mass effects. At this point, however, we are not primarily interested in the differences of the pairing gaps \emph{between} the different classes of Table~\ref{tab:eps_classes}, but rather in the deviation of the pairing gaps \emph{within} a given class as we approximate the momentum dependence of the normal self-energy and of the $Z$-factor.

It is clear from Fig.~\ref{fig:gap_hard} that the choice of the method used to average the momentum dependence of the effective mass and of the $Z$-factor has a strong impact on pairing gaps, irrespective if $e$-mass effects are taken into account or not. This result could have been expected from Fig.~\ref{fig:mkf_SNM_BHF}. In contrast to the soft interaction case (lower panel of Fig.~\ref{fig:gap_soft} for SNM), replacing $\varepsilon^{\rm{BHF}} (p) \rightarrow (p^2 - \kf^2)/(2  m^{* \rm{BHF}} (\kf))$  is unreliable. Since $m^{*\rm{BHF}}_{av}$ is systematically larger than $m^{*\rm{BHF}}_{pe}$ (see Fig.~\ref{fig:mkf_SNM_BHF}), we find for hard potentials, in contrast to Eq.~(\ref{eq:gaps_HF}),
\begin{equation}
\widehat{\Delta} \big[ m^{*\rm{BHF}}_{av} (\kf) \big] > \widehat{\Delta} \big[ \varepsilon^{*\rm{BHF}} (k) \big] > \widehat{\Delta} \big[ m^{*\rm{BHF}}_{pe} (\kf) \big] \, ,
\end{equation}
for all densities of interest. Such an inequality holds irrespective if $e$-mass effects are taken into account or not. We also observe that the pairing gaps obtained for the two classes are bounded from above, over the entire density range, by the lower limit of the band obtained when using the single-particle spectrum generated by the SLy4 Skyrme functional and without explicit $Z$-factor.

\subsubsection{Analysis}
\label{results_gaps_analysis}

\begin{figure}
\includegraphics[clip]{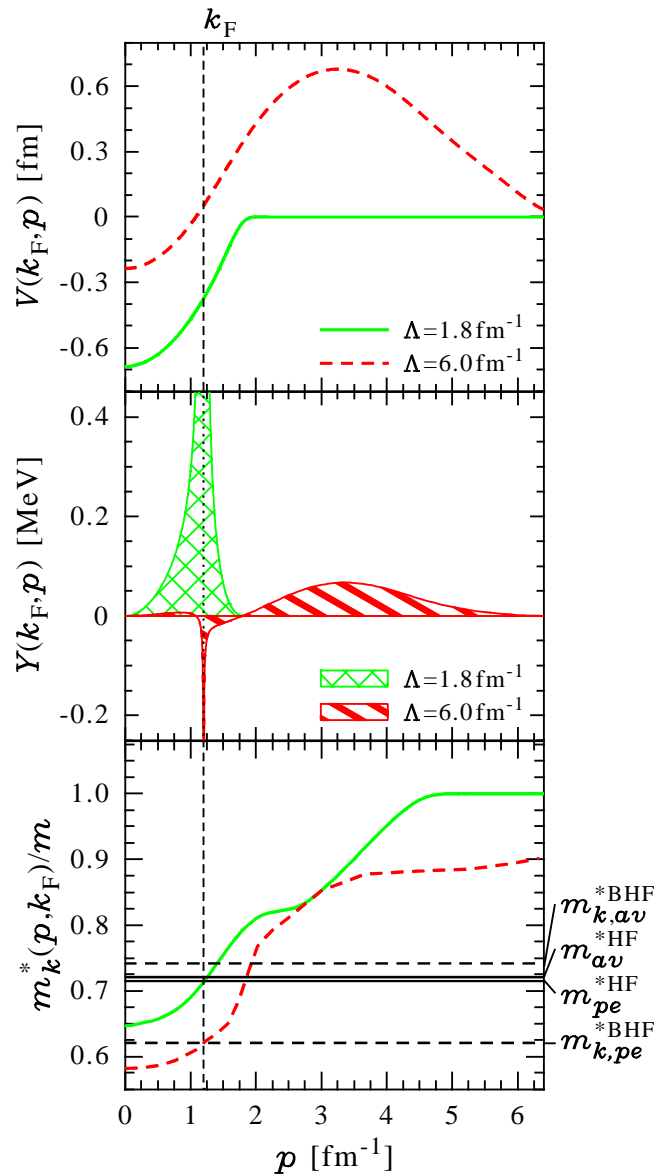}
\caption{(Color online) Analysis of the effective-mass approximation at a representative density $\kf = 1.2\,\rm{fm}^{-1}$. Middle panel: (i) for low cutoffs the gap at the Fermi momentum is built from low-momentum modes around the Fermi surface, and (ii) for large cutoffs the gap is built mainly from high-momentum modes, where the NN interaction matrix elements are maximal and repulsive (upper panel). In the hard potential case, the momentum averaging of the effective mass would necessitate a non-trivial fine tuning to reproduce the pairing gaps calculated with the full momentum dependence of the $k$-mass (lower panel).}
\label{fig:gap_analysis}
\end{figure}

In order to get a deeper understanding of the qualitative difference between soft and hard interactions, we write the gap equation, Eq.~(\ref{eq:gap}), as
\begin{equation}
\widehat{\Delta} (p=\kf) \equiv \int dq \, Y(\kf, q) \, .
\end{equation}
Once the self-consistent gap equation is solved, the integrand function $Y(\kf, p)$ contains information about the momentum scales from which the gap at the Fermi momentum is built. The NN interaction matrix elements $V(\kf, p)$, the resulting function $Y(\kf, p)$, and the $k$-mass are shown in Fig.~\ref{fig:gap_analysis} for cutoffs $\Lambda_{\text{soft}}=1.8\,\rm{fm}^{-1}$ and $\Lambda_{\text{hard}}=6.0\,\rm{fm}^{-1}$ at a representative density of $\kf = 1.2\,\rm{fm}^{-1}$. For simplicity, we only consider the class $\Lambda^1_{\rm{hard}}$ for the hard interaction. This way, we avoid the additional subtleties connected with $Z$-factors and the more pronounced momentum dependence of the total effective mass for the hard potential case.

As can be seen from the middle panel of Fig.~\ref{fig:gap_analysis}, the gap is generated for soft interactions from momentum modes around the Fermi surface since off-diagonal matrix elements do not couple low and high momenta (upper panel). It is therefore understandable that fixing the momentum-independent effective mass to its value at or in the vicinity of the Fermi surface (lower panel) is a good approximation. For large cutoffs, however, the major contributions to the gap at the Fermi momentum originate from high momenta~\cite{Baldo90}, far away from the Fermi surface. In particular for $\Lambda=6.0\,\rm{fm}^{-1}$ and $\kf = 1.2\,\rm{fm}^{-1}$, the matrix elements of the pairing interaction are very small around the Fermi surface and essentially the entire gap strength is built from the repulsive part of the NN interaction at high momenta. Therefore, approximating the normal self-energy through a momentum-independent effective mass $m^{*\rm{BHF}}_{pe/av}$ defined in the vicinity of the Fermi surface is unreliable in this case. Since the effective mass approaches the free mass $m$ at high momenta (see lower panel), the gaps $\Delta [m^{*\rm{BHF}}_{pe}]$ are too small compared to the reference ones obtained by keeping the full momentum dependence of the normal self-energy. In contrast, the gaps $\Delta[m^{*\rm{BHF}}_{av}]$ are too large compared to the reference gaps since the averaging method, Eq.~(\ref{eq:av}), includes high-momentum modes with too much weight, where the effective mass is larger than at the Fermi surface. Hence, in order to obtain a more reliable momentum-independent effective mass for hard potentials, the behavior of the NN interaction matrix elements would have to be taken into account in the averaging scheme.

\section{Finite nuclei results}
\label{fin_nuclei}

Before going to higher orders, our goal is to perform a complete first-order calculation in superfluid nuclei, including the normal and anomalous self-energies consistently. Intermediate steps towards this goal have been taken recently. On the one hand, calculations of non-superfluid nuclei have been performed starting from soft NN interactions~\cite{coraggio03a,roth06a}. This gives access to the single-particle field at first order in the NN interaction but does not account for the pairing channel. Recently, the work of Ref.~\cite{roth06a} has been extended to HFB calculations of superfluid nuclei, employing a soft NN interaction in the particle-hole channel but without three-nucleon (3N) forces~\cite{Hergert:2009nu}. On the other hand, HFB calculations combining a pairing kernel based on the direct NN interaction with the single-particle field generated by the empirical SLy4 Skyrme functional have been performed~\cite{Duguet07,Lesinski08,Barranco04,pastorethesis,Pastore08}. Eventually, one needs to combine the benefits of these two applications to obtain complete first-order calculations of superfluid nuclei, including also the effects of 3N interactions on the single-particle field. Although the density of states around the Fermi energy may be dominated by NN interactions, 3N interactions contribute especially to spin-orbit splittings in nuclei. The present work is an attempt to qualify whether or not the two sets of results published in Refs.~\cite{Duguet07,Lesinski08} and Refs.~\cite{Barranco04, Pastore08} provide a good approximation to a complete first-order calculation and to understand the mismatch between them. While insights have been obtained in previous sections through calculations in INM, this present section is dedicated to assessing the situation in finite nuclei.

\subsection{Soft interaction}
\label{finite_soft}

\begin{figure}
\includegraphics[clip,width=8.5cm]{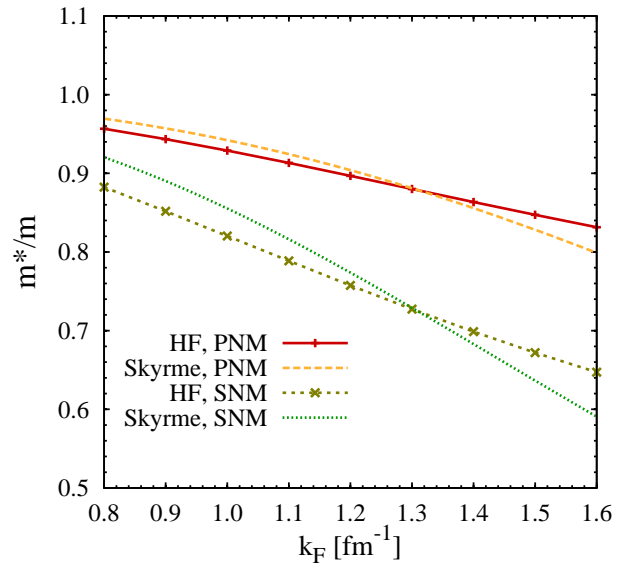}
\caption{(Color online) New Skyrme effective masses constrained to reproduce the HF result obtained from the soft NN interaction.}
\label{fig:refittedmasses1}
\end{figure}

\begin{figure}
\includegraphics[width=8.5cm,clip]{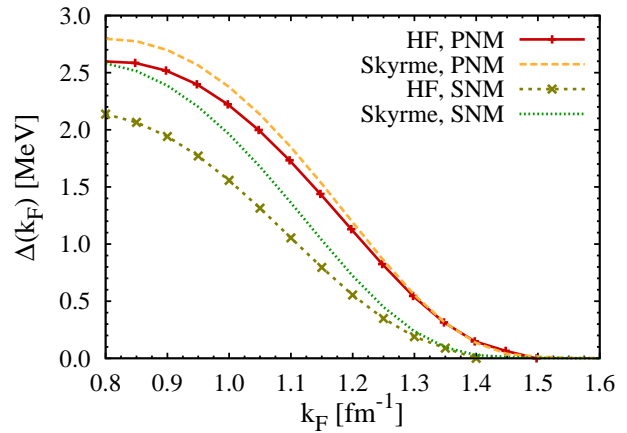}
\caption{(Color online) Neutron $^1$S$_0$ pairing gaps in SNM and PNM based on the refitted Skyrme EDF (see text) with the soft NN interaction as the pairing kernel.}
\label{fig:recalculatedgaps0}
\end{figure}

Let us start with the case of the soft NN interaction. The findings discussed in Section~\ref{results_gaps_soft} are useful to assess the validity of the results of Refs.~\cite{Duguet07,Lesinski08}. One expects the momentum-independent effective-mass approximation, which is reliable in INM, to be tractable in finite nuclei as well, because the discussion of Section~\ref{results_gaps_analysis} can be carried over to finite nuclei. This gives some confidence that computing the single-particle field from an EDF characterized by a momentum-independent effective mass is a good approximation, as long as one is working consistently with a low resolution scale. Nevertheless, this must be checked explicitly by comparing the results thus obtained with the HF single-particle field computed in finite nuclei\footnote{References~\cite{coraggio03a,roth06a} focus on the single-particle spectrum of $^{40}$Ca, which displays significant differences with the corresponding spectrum generated from Skyrme EDFs characterized by $m^{\ast}/m=0.7$ at nuclear saturation density. However, $^{40}$Ca constitutes an anomaly, because the spectrum generated by such EDFs is unnaturally dense around the Fermi energy~\cite{Lesinski07,brown98a}. In addition, a meaningful microscopic calculation of single-particle spectra must include 3N interactions.} from soft NN~\cite{coraggio03a,roth06a} and 3N interactions.

\begin{figure*}
\includegraphics[width=16cm, clip]{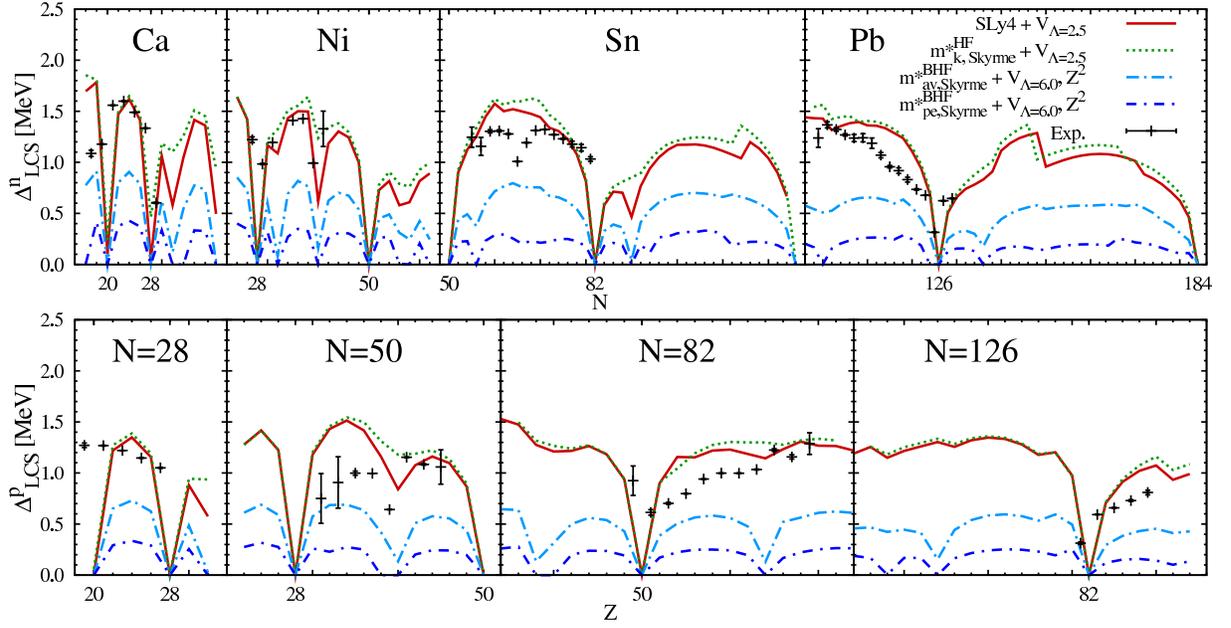}
\caption{(Color online) Neutron and proton LCS pairing gaps computed in semi-magic nuclei using a soft/hard NN interaction and the corresponding refitted Skyrme EDF whose effective masses are shown in Fig.~\ref{fig:refittedmasses1}/Fig.~\ref{fig:refittedmasses2}. In the hard-interaction case, a $Z$-factor is taken into account when solving the HFB equations. Calculations include the Coulomb interaction in the proton pairing kernel~\cite{Lesinski08}.}
\label{recalculatedgaps1}
\end{figure*}

\begin{figure*}
\includegraphics[width=16cm, clip]{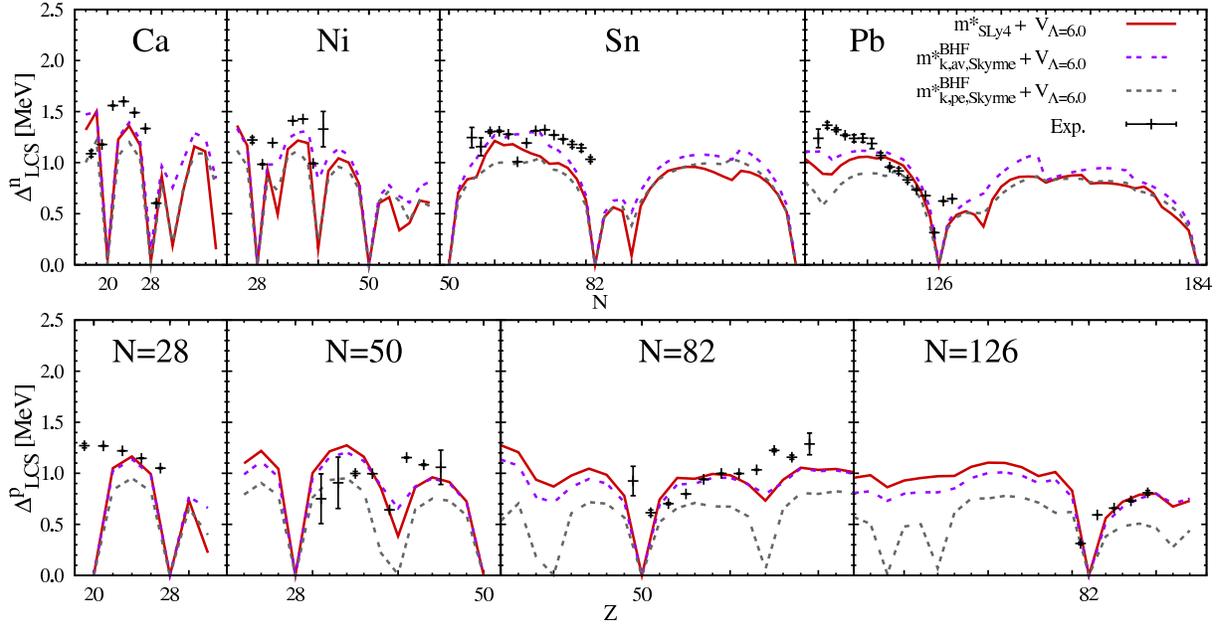}
\caption{(Color online) Neutron and proton LCS pairing gaps computed in semi-magic nuclei using a hard NN interaction and the corresponding refitted Skyrme EDF whose effective masses are shown in Fig.~\ref{fig:refittedmasses3}. The $Z$-factor and $e$-mass are not taken into account when solving the HFB equations. Calculations include the Coulomb interaction in the proton pairing kernel~\cite{Lesinski08}.}
\label{recalculatedgaps2}
\end{figure*}

\subsubsection{Refitted Skyrme effective mass}

Given the validity of the momentum-independent effective-mass approximation, we construct a new empirical EDF so that it reproduces well the effective mass obtained microscopically. As discussed, to this end the SLy4 value $m^{\ast}/m=0.7$ at nuclear saturation density is appropriate. However, the SLy4 effective mass has an incorrect isovector dependence and an unsatisfactory low-density behavior in SNM (see Fig.~\ref{fig:mkf_HF_LSy4}). To improve on these deficiencies and to reach a higher confidence in the gaps of Refs.~\cite{Duguet07,Lesinski08}, we generate a new parameterization of the Skyrme EDF, adding to the Lyon protocol~\cite{chabanat98} the constraint that both SNM and PNM HF effective masses shown in Fig.~\ref{fig:mkf_HF_LSy4} are reproduced around saturation density. The refitted effective masses in both SNM and PNM are shown in Fig.~\ref{fig:refittedmasses1}. By default, the isoscalar effective mass reproduces well the microscopic one around saturation, whereas the isovector one is clearly improved compared to SLy4. However, the isoscalar effective mass is only marginally improved compared to SLy4 at low densities where it overestimates the HF result. In fact, a better reproduction of the low-density behavior of the effective mass in SNM requires extending the analytical form of the Skyrme functional~\cite{lesinski06a}. This is underway, in particular through the design of non-empirical Skyrme EDFs based on the development of the density-matrix expansion for low-momentum interactions~\cite{Bogner:2008kj,gebremariam09a}.

Pairing gaps calculated in SNM and PNM with the refitted Skyrme EDF and the soft NN interaction as a pairing kernel are presented in Fig.~\ref{fig:recalculatedgaps0}. With the improved isovector effective mass the gaps are more satisfactory in PNM compared to using SLy4 (see upper panel of Fig.~\ref{fig:gap_soft}), but they are essentially unchanged in SNM, where in particular, microscopic pairing gaps are overestimated for $\kf < 1.2\,\rm{fm}^{-1}$.

\subsubsection{Pairing gaps in semi-magic nuclei}

Employing the refitted Skyrme EDF and the soft NN interaction as the pairing kernel, we compute neutron and proton pairing gaps in semi-magic nuclei by solving HFB equations in spherical symmetry~\cite{Lesinski08}. Theoretical gaps are provided by
$\Delta_{\rm LCS}$ which denotes the diagonal pairing matrix
element $\Delta_i$ corresponding to the \emph{canonical} single-particle state
$\phi_i$ whose quasi-particle energy\footnote{The acronym LCS stands for Lowest Canonical State.}
is the lowest. Our results are presented in Fig.~\ref{recalculatedgaps1}. Experimental gaps extracted from binding energies through three-point mass differences centered on odd-mass nuclei~\cite{Duguet01b} are shown as a reference. The pairing gaps are essentially identical to those obtained using SLy4~\cite{Lesinski08} where data exist. The improved isovector effective mass leads to a tiny increase (decrease) of neutron (proton) gaps in neutron-rich nuclei. Note that, although the discrepancy between the refitted Skyrme effective mass and the microscopic results at low densities in SNM implies an uncertainty for the present results, we expect this to be small, as it is unlikely that such densities weigh significantly in pairing gaps of (non-halo) nuclei.

Keeping in mind the necessity to confirm the effective-mass approximation through a systematic comparison of HF and Skyrme single-particle spectra in doubly-magic nuclei, one can conclude that the NN-only results of Refs.~\cite{Duguet07,Lesinski08} are presently put on a rather solid basis. The most important of these conclusions is that neutron and proton pairing gaps in semi-magic spherical nuclei are approximately accounted for using the $^1$S$_0$ partial wave of soft interactions at first order. This result is valid~\cite{Lesinski08} over the broad cutoff range of $\Lambda\approx1.8-3.0\,\rm{fm}^{-1}$ that characterizes perturbative NN interactions~\cite{Bogner:2006tw,Bogner05,Bogner:2009un}. This finding is somewhat puzzling as it indicates that neglected contributions, such as higher partial waves, 3N interactions and coupling to density, spin and isospin collective fluctuations for both the normal self-energy and the pairing interaction, may have a small net effect on pairing gaps in nuclei\footnote{Of course, the neglected contributions do not have to be individually small.}. Such a conjecture needs to be confirmed by incorporating explicitly all neglected contributions in a consistent way.

\subsection{Hard interaction}
\label{finite_hard}

\begin{figure}[t]
\includegraphics[width=8.5cm, clip]{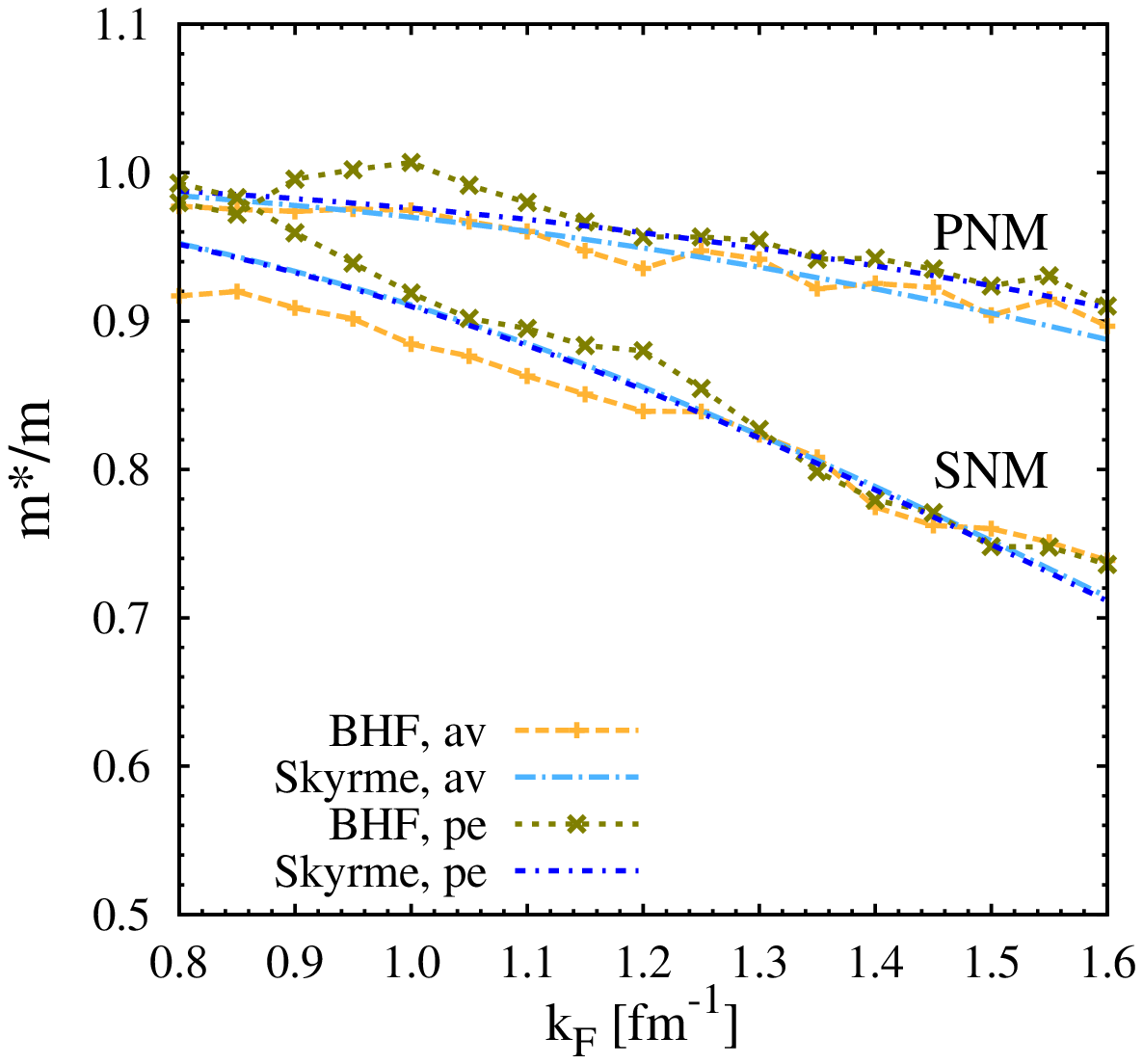} \vspace{0.3cm} \\
\includegraphics[width=8.5cm, clip]{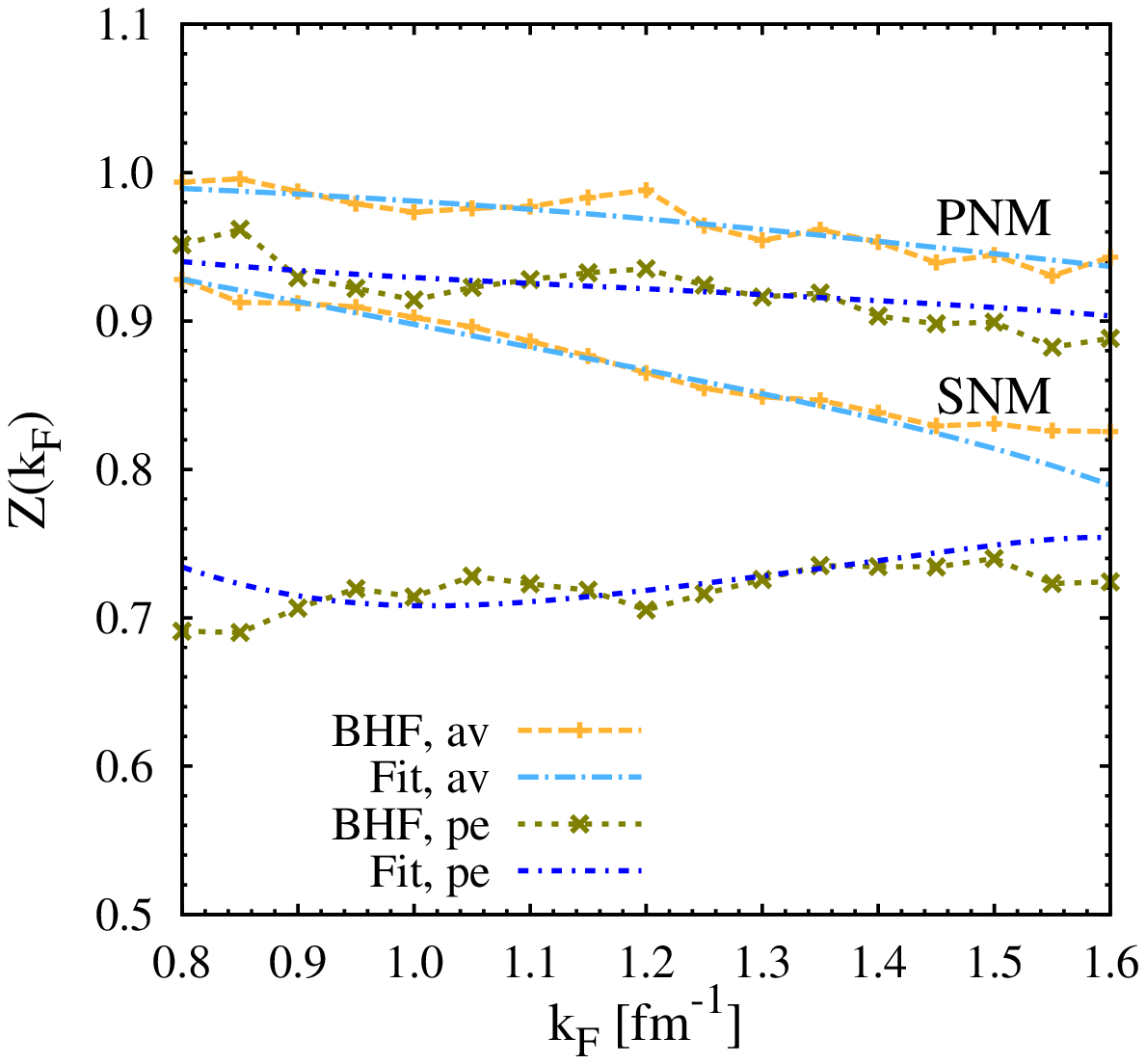}
\caption{(Color online) New Skyrme effective masses constrained to reproduce BHF total effective masses obtained from the hard NN interaction. We have used a simple functional form to fit the $Z$-factors.}
\label{fig:refittedmasses2}
\end{figure}

Let us now turn to the hard NN interaction case. The findings discussed in Section~\ref{results_gaps_hard} are useful to assess the validity of the results of Refs.~\cite{Barranco04,Pastore08}. Indeed, it is natural to expect that the uncertainties in  pairing gaps generated in INM by the momentum-independent effective-mass approximation propagate to finite nuclei. This is the case because the uncertainty relates to high-momentum modes that are of a similar nature in homogeneous and non-homogenous systems. Practical applications combining a hard NN interaction as pairing kernel with a momentum-independent effective mass provided by a phenomenological Skyrme functional would require a fine tuning of the momentum-independent effective mass and of the momentum-independent $Z$-factor when solving HFB equations. Due to the energy dependence of the BHF normal self-energy, a consistent calculation of pairing gaps within the Brueckner expansion requires the inclusion of $Z$-factors. However, since the HFB calculations of Refs.~\cite{Barranco04,Pastore08} are performed without $Z$-factor, both classes $\Lambda^Z_{\rm{hard}}$ and $\Lambda^1_{\rm{hard}}$ (see Section~\ref{gaps}) are investigated in this study in order to make contact with these works.

\subsubsection{Refitted Skyrme effective mass and $Z$-factor}

In order to assess the uncertainties of the pairing gaps in finite nuclei due to the effective-mass approximation schemes, we use the INM results of Section \ref{results_effmassavermomentumdep} for the construction of new Skyrme functionals for hard interactions. We again consider the two classes of Table~\ref{tab:eps_classes}: $\Lambda^Z_{\rm{hard}}$ with $m^* = m^*_{pe/av}$ and $Z = Z_{pe/av}$ as well as $\Lambda^1_{\rm{hard}}$ with $m^* = m^*_{k,pe/av}$ and $Z = 1$. For the two classes, the difference of the gaps using the point-evaluated and averaged quantities provides a range for the uncertainty in the pairing gaps.

\begin{figure}
\includegraphics[width=8.5cm, clip]{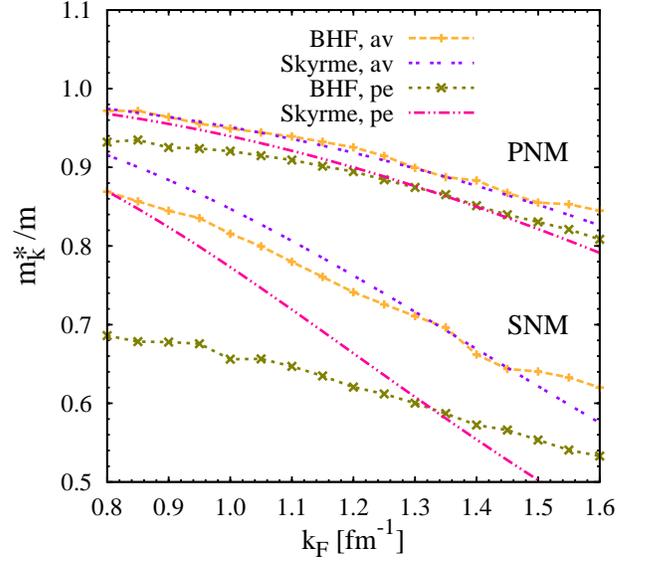}
\caption{(Color online) New Skyrme effective masses constrained to reproduce BHF $k$-masses obtained from the hard NN interaction.}
\label{fig:refittedmasses3}
\end{figure}

The refitted Skyrme effective masses in SNM and PNM are shown in Figs.~\ref{fig:refittedmasses2} and~\ref{fig:refittedmasses3}. As in the soft-interaction case, the effective mass is satisfactory around saturation density in both SNM and PNM. Only the fit to the point-evaluated $k$-mass in SNM is somewhat problematic due to the density dependence of the Skyrme parametrization. Reproducing well the BHF effective mass at low densities would require to extend the analytical form of the Skyrme functional~\cite{lesinski06a}.

\subsubsection{Pairing gaps in semi-magic nuclei}
\label{recalcgapsfinitenucleihard}
Employing the refitted Skyrme EDFs and the hard NN interaction as the pairing kernel, we compute neutron and proton pairing gaps in semi-magic nuclei. The results are shown in Figs.~\ref{recalculatedgaps1} and \ref{recalculatedgaps2}.

We observe a reduction of the pairing gaps obtained within the $\Lambda^Z_{\rm{hard}}$ class compared to those obtained from soft NN interactions. This is consistent with the results obtained in INM and discussed in Section~\ref{results_gaps}. This difference is genuine and reflects that many-body expansion schemes depend on the resolution scale and that finite-order results for soft and hard interactions are in general not equivalent and immediately comparable. Effectively, the $Z$-factor accounting for the energy dependence of the BHF self-energy is largely responsible for the smallness of the pairing gaps computed at lowest order in the hard NN interaction. Omitting the $Z$-factor and $e$-mass makes the present calculation formally similar to those performed in Refs.~\cite{Barranco04,Pastore08} which (using the SLy4 parameterization with an isoscalar effective mass of $m^*/m=0.7$ at saturation density) led to pairing gaps smaller by a factor of two-thirds compared to those obtained with soft interactions~\cite{Lesinski08}. Using the new Skyrme parametrization, such a calculation leads to larger gaps than in class $\Lambda^Z_{\rm{hard}}$, but still generally smaller than with soft interactions. However, for both classes $\Lambda^Z_{\rm{hard}}$ and $\Lambda^1_{\rm{hard}}$, the dependence of the pairing gaps on the effective-mass approximation scheme is significant. For class $\Lambda^Z_{\rm{hard}}$ the uncertainty is $\approx0.5\,\rm{MeV}$ and for class  $\Lambda^1_{\rm{hard}}$ it is $\approx0.25\,\rm{MeV}$, which both constitute a substantial fraction of the gap strength. This demonstrates that the effective-mass approximation problem is present irrespective of the inclusion or omission of the energy dependence of the self-energy.

\section{Conclusions}
\label{conclusions}
The present paper complements recent works directed towards the
construction of non-empirical energy functionals for
nuclei~\cite{Barranco04,Pastore08,Duguet07,Lesinski08,pastorethesis}.
We have studied neutron $^1$S$_0$ pairing gaps with special attention
to the consistency of the pairing interaction and normal self-energy
contributions. In nuclear matter, we calculated the normal and
anomalous parts of the self-energy consistently at first order in the
expansion scheme for soft and hard NN interactions. Our results also
provide new constraints to empirical Skyrme functionals. We have found
that $T=1$ pairing gaps obtained from low-momentum interactions depend only weakly on approximations to the normal self-energy, while gaps from hard potentials are very sensitive to the effective-mass approximation scheme. This is because a momentum-independent effective mass does not approach the free mass at high momenta, but for hard interactions the high-momentum modes are not decoupled. The same conclusion has been reached for calculations of pairing gaps in finite nuclei. This is problematic for hard NN interactions when employed in conjunction with standard empirical EDFs which are of low-momentum character. Although a complete first order calculation is needed, where the Skyrme EDF is replaced by a microscopic HF calculation including 3N forces, our results put the effective-mass approximation used in Refs.~\cite{Duguet07,Lesinski08} on a rather solid basis.

\section{Acknowledgments}

We thank S.\ Baroni, F.\ Barranco, P.\ F.\ Bortignon, R.\ A.\ Broglia, A.\ Pastore and E.\ Vigezzi for useful discussions. This work was supported in part by the Natural Sciences and Engineering
Research Council of Canada (NSERC), the U.S.~Department of Energy  under Contract Nos. DE-FG02-96ER40963, DE-FG02-07ER41529 (University  of Tennessee) and DE-AC05-00OR22725 with UT-Battelle, LLC (Oak Ridge  National Laboratory). TRIUMF receives federal funding via a contribution agreement through the National Research Council of Canada. 

\bibliography{pairing}

\end{document}